\documentclass{aa}  
\usepackage{graphicx}
\usepackage{txfonts}
\def\simlt{\lower.5ex\hbox{$\; \buildrel < \over \sim \;$}}
\def\simgt{\lower.5ex\hbox{$\; \buildrel > \over \sim \;$}}
\def\ltsim{\raise 2pt \hbox {$<$} \kern-1.1em \lower 4pt \hbox {$\sim$}}
\def\ltapprox{\raise 2pt \hbox {$<$} \kern-1.1em \lower 5pt \hbox {$\approx
$}}
\def\gtsim{\raise 2pt \hbox {$>$} \kern-1.1em \lower 4pt \hbox {$\sim$}}
\def\gtapprox{\raise 2pt \hbox {$>$} \kern-1.1em \lower 5pt \hbox {$\approx
$}}
\begin{document}

\title{Brightest Cluster Galaxies in the Extended GMRT radio halo cluster 
sample}
\subtitle{Radio properties and cluster dynamics}
\author{R.~Kale\inst{1,2}, T. Venturi\inst{1}, R. Cassano\inst{1}, S. Giacintucci\inst{3}, S. Bardelli\inst{4}, D.Dallacasa\inst{5,1}, E. Zucca\inst{4}}

\institute
{INAF--Istituto di Radioastronomia, Via Gobetti 101, 40129 Bologna, Italy\\
\email{tventuri@ira.inaf.it}
\and
National Centre for Radio Astrophysics, Tata Institute of Fundamental 
Research, Post Bag 3, Ganeshkind, Pune 411007, India
\and
Department of Astronomy, University of Maryland, College Park, MD, 20742-2421, 
USA
\and
INAF--Osservatorio Astronomico, Via Ranzani 1, 40126 Bologna, Italy
\and
Dipartimento di Fisica e Astronomia, Universit\'a di Bologna, Via Ranzani 1,
40126 Bologna, Italy
}

\date{Received 00 - 00 - 0000; accepted 00 - 00 - 0000}

\titlerunning{Radio properties of BCGs in the Extended GMRT cluster sample}
\authorrunning{Kale et al.}


\abstract
{}
{First-ranked galaxies in clusters, usually referred to as Brightest 
Cluster Galaxies (BCGs), show exceptional properties over the whole 
electromagnetic spectrum. 
They are the most massive elliptical galaxies, and show the 
highest probability to be radio loud. Moreover, their special location at 
the centres of galaxy clusters raises the question of the role of 
the environment on their radio properties.  
In the attempt to decouple the effect of the galaxy mass and of the 
environment in their statistical radio properties, we investigate the possible 
dependence of the occurrence of radio loudness and of the fractional
radio luminosity function on the dynamical state of the hosting cluster.}
{We studied the radio properties of the BCGs in the Extended GMRT Radio 
Halo Survey (EGRHS), which consists of 65 clusters in the redshift range 
0.2--0.4,
with X--ray luminosity L$_{\rm X} \ge 5\times 10^{44}$ erg~s$^{-1}$,
and quantitative information on their dynamical state from high quality 
$Chandra$ imaging. We obtained a statistical sample of 59 BCGs, which
was divided into two classes, depending on the dynamical state of the host 
cluster, i.e. merging (M) and relaxed (R).}
{Among the 59 BCGs, 28 are radio--loud, and 31 are radio--quiet. The 
radio--loud sources are located favourably located in relaxed clusters (71\%), 
while the reverse is true for the radio--quiet BCGs, mostly located in merging 
systems (81\%). 
The fractional radio luminosity function (RLF) for the BCGs 
in merging and relaxed clusters is different, and it is considerably higher 
for BCGs in relaxed clusters, where the total fraction of radio loudness 
reaches almost 90\%, to be compared to the $\sim$30\% in merging clusters.
For relaxed clusters, we found a positive correlation between the radio power 
of the BCGs and the strength of the cool core, consistent with previous
studies on local samples. }
{Our study suggests that the radio loudness of the BCGs strongly depends 
on the cluster dynamics, their fraction being considerably higher in relaxed 
clusters. We compared our results with similar investigations, and briefly 
discussed them in the framework of AGN feedback.}

\maketitle
\section{Introduction}\label{sec:intro}
First-ranked galaxies are the brightest and most massive galaxies
in the Universe, and inhabit the cores of galaxy clusters. 
Galaxies in this class are both elliptical and cD, and
are commonly referred to as Brightest Cluster Galaxies (BCGs).
BCGs represent the bright end of the luminosity function of early--type 
galaxies, with a small luminosity dispersion around the mean value. 
They are located at small distance from the peak of the thermal X--ray 
emission from the intracluster medium (ICM), and have small velocity 
dispersions (Quintana \& Lawrie \cite{quintana82}). 

Due to their special location at the centres of the largest gravitationally 
bound structures in the Universe, BCGs have been devoted special
attention for a long time. Many of them exhibit exceptional properties, 
with emission in the UV and FIR, as well as H$_{\alpha}$ lines, suggesting 
the presence of multiphase gas and ongoing star formation 
(e.g. O'Dea et al. \cite{odea08}, Haarsma et al. \cite{haarsma10}, 
Donahue et al. \cite{donahue10}, Edge et al. \cite{edge10}, 
O'Dea et al. \cite{odea10}, Liu et al. \cite{liu12}).
\\
From the radio point of view, BCGs are a special class, too. A large 
fraction of them shows radio emission of nuclear origin. In a number of cases 
the radio emission extends well beyond the optical envelope to form extended
radio jets, which bend in a C shape (wide--angle tail sources) as a result 
of galaxy motion and cluster weather (Burns \cite{burns98}). A prototypical
case is the radio galaxy 3C\,465 at the centre of A\,2634 (Eilek et al. 
\cite{eilek84}). Albeit some
remarkable exceptions, their radio power is either at the transition 
between FRI and FRII radio galaxies 
(Fanaroff \& Riley \cite{fr74}) or below (Owen \& Laing \cite{ol89}).

Over the past decade, our view and understanding of the properties of the
central regions in galaxy clusters 
has improved thanks to the contribution of the X--ray observatories 
$Chandra$ and {\it XMM--Newton}.
The radiative cooling of the X--ray emitting gas in cool--core clusters 
(Peterson \& Fabian \cite{peterson06}) requires some source of heating to 
balance the radiative losses, and the AGN activity associated with 
the BCG in those systems is the primary candidate to provide such energy
(McNamara \& Nulsen \cite{mcnamara07}). The existence of aged radio plasma, 
detected at frequencies below 1 GHz and associated with X--ray cavities in 
a number of rich and poor clusters, is interpreted as the signature of 
repeated radio outbursts from the BCG, and provides strong support to the 
AGN feedback picture (e.g., Clarke et al. \cite{clarke05} and \cite{clarke09}, 
Fabian et al. \cite{fabian02}, Giacintucci et al. \cite{giacintucci11a},
McNamara \& Nulsen \cite{mcnamara12}).

An important tool to investigate the nature of the radio loudness in 
elliptical galaxies is the fractional radio luminosity function (RLF), 
defined as the probability that an elliptical galaxy of a given optical 
magnitude hosts a radio galaxy with radio power above a threshold value. 
A number of studies show that the RLF strongly depends on the optical 
magnitude of the associated galaxy (e.g. Auriemma et al. \cite{auriemma77},
Ledlow \& Owen \cite{lo96}, Mauch \& Sadler \cite{mauch07},
Bardelli et al. \cite{bardelli10}), and is higher
for brighter absolute optical magnitudes. BCGs are by definition the 
brightest galaxies, and show the highest probability to be radio loud.
However, BCGs are special not only in terms of mass but also because of 
their location at the cluster centre, and it is important to decouple
these two effects.
Those earlier studies suggested that the fractional radio luminosity
function is independent of the galaxy environment (rich clusters, groups,
field). On the other hand, Best et al. (\cite{best07}) found that 
BCGs are more likely to be radio loud than other galaxies of similar
mass, and this effect becomes stronger for galaxies with stellar
mass M$<10^{11}$M$_{\rm Sun}$, suggesting that their location at the cluster 
centre does play a role in their radio properties.
The importance of the local environment was clear also from the work of
Mittal et al. (\cite{mittal09}), who studied the radio properties of the 
HIghest X--ray FLUx Galaxy Cluster Sample (HIFLUGCS, 
Reiprich \& B\"ohringer \cite{RB02}), and found that radio loud BCGs are 
more abundant in cool--core clusters: their fraction increases from 45\%
in non--cool--core (NCC) to 100\% in strong cool--core (SCC) systems,
and their radio power shows a positive correlation with the cool--core
strength.

In this context, and motivated by the importance to decouple the effect 
of the galaxy mass from that of the local environment, we addressed
the question of the radio properties of BCGs in connection with the 
dynamical status of the host cluster, with the aim of providing a 
complementary picture to previous literature studies. 
We used the Extended GMRT (Giant Metrewave Radio Telescope) Radio Halo Survey 
(EGRHS), which includes 65 clusters in the redshift interval 0.2$\le z \le$0.4 
observed at 610 MHz 
(see Venturi et al. \cite{venturi07} and \cite{venturi08}, hereinafter V07 
and V08; Kale et al. \cite{kale13} and \cite{kale15}, hereinafter K13 and 
K15).
The main goal of the EGRHS was to investigate the origin of diffuse 
cluster--scale radio sources in galaxy clusters, namely radio halos, 
mini--halos and relics. Thanks also to the results of the EGRHS, nowadays
it is quite clear that the origin of large diffuse emission
in galaxy clusters is related to the cluster dynamical status 
(e.g. Brunetti \& Jones \cite{bj14} for a review). In particular, it
has been statistically shown that giant radio halos are associated with
merging galaxy clusters (V07, Cassano et al. \cite{cassano10}, hereinafter
C10), as well as radio relics (e.g. de Gasperin et al. \cite{degasperin14}).
On the other hand, radio mini--halos always surround a radio active BCG
at the centre of relaxed cool--core clusters (Giacintucci et al.
\cite{giacintucci14}, ZuHone et al. \cite{zuhone13}, K15). 
Diffuse cluster scale emission in galaxy clusters can thus be used as
tracer of the cluster dynamics, together with the more direct probes
supplied by X--ray imaging and analysis.

In this paper we present the radio properties and the fractional radio
luminosity function of the BCGs in the EGRHS, and relate those quantities
to the cluster dynamical status (merger versus relaxed), which we
derived quantitatively using high quality $Chandra$ images.
The paper is organized as follows. In Sect. \ref{sec:sample} we report on 
the selection criteria for our BCG sample and provide an overview of the 
sample as a whole; the radio properties of the BCGs and the X--ray 
properties of the host clusters are presented in in Sect.  \ref{sec:radioX}; 
in Sect. \ref{sec:analysis} we present the statistical radio properties of
the BCGs, as well  as the dynamical properties of the host cluster, and we 
describe the method to derive the fractional radio luminosity function.
A discussion of our results and conclusions are given respectively
in Sect. \ref{sec:disc} and \ref{sec:conc}.

We adopted a standard $\Lambda$ CDM cosmology to convert observed
quantitites into intrinsic ones (H$_{\rm o}$=70 km~s$^{-1}$Mpc$^{-1}$, 
$\Omega_{\rm M}$=0.29). The convention S$\propto\nu^{-\alpha}$ is used
throughout the paper.


\section{BCG Sample}\label{sec:sample}

The sample of BCGs presented in this work is extracted from the EGRHS
(V07, V08, K13 and K15), 
which consists of galaxy clusters 
selected from the ROSAT--ESO flux--limited X--ray (REFLEX) galaxy cluster 
catalogue  (B\"ohringer et al. \cite{boh04}) and from the extended ROSAT 
Brightest Cluster Sample (EBCS) catalogue (Ebeling et al. \cite{ebeling98} and 
\cite{ebeling00}) according to the following criteria:
\begin{itemize}
 \item{} L$_X$ (0.1-2.4 keV) $> 5\times10^{44}$ erg s$^{-1}$;
 \item{} $0.2 \le z \le 0.4 $; and
 \item{} $\delta  > -31^{\circ}$.
\end{itemize}
From the original cluster sample
we removed A\,689, whose
X--ray luminosity has been recently revised and is below our threshold  
(Giles et al. \cite{giles12}), and we remained with 65 clusters, whose BCGs 
were identified by visual inspection of the optical images. Where available 
we used images from the Data Release 7 (DR7) of the Sloan Digital Sky Survey 
(SDSS, Ahn et al. \cite{ahn14}), otherwise we used the red plate of the
Digitized Sky Survey (DSS-2).
In order to identify the BCGs we searched for the brightest cluster member 
in the proximity of the X--ray surface brightness peak, using the NASA
Extragalactic Database (NED).
Proprietary and archival $Chandra$ and $XMM-Newton$ X--ray images were 
used to this aim. 
Out of the full cluster sample, six clusters host 2 BCGs (see Sect. 2.1),
while no obvious one was found in three clusters (see Sect. 2.2).

The final BCG sample includes 68 objects (in 62 clusters). 
Table \ref{tab:id} reports the galaxies used for our statistical
analysis, while Table \ref{tab:id2} lists those radio emitting BCGs
which were excluded (see Sect. 3.1). Both tables are listed in
order of decreasing radio power, and contain the following
information: column 1=cluster name; column 2= redshift; column 3=name 
of the BCG (from NED); column 4=radio power at 1.4 GHz (see Sect. 3.1); 
column 5=note on the dynamical state of the cluster (see Sect. 3.2); 
column 6 = note on the presence of diffuse cluster scale emission 
(RH=radio halo, MH=mini--halo). 
The three clusters with no obvious BCG are listed in Table \ref{tab:noBCG}. 
Some notes on the special cases are given in the next subsections. 

\begin{figure}[htbp]
\includegraphics[angle=0,scale=0.48]{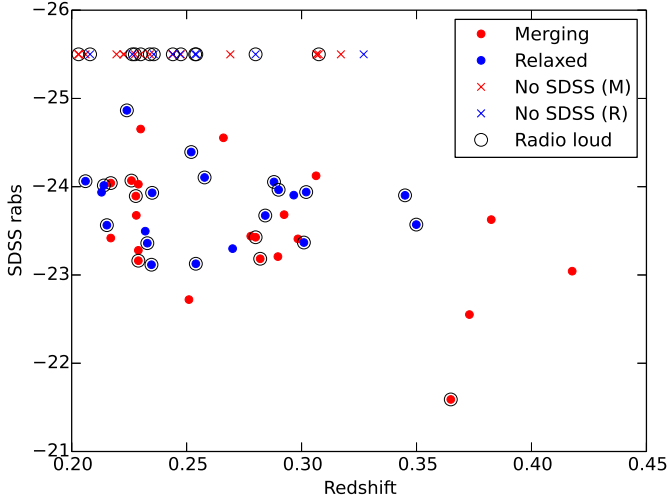}
\caption{Distribution of the SDSS absolute red magnitude of the BCGs in 
the sample. Red and blue dots show the BCGs in merging and relaxed clusters
respectively (see Sect. 3.2). Crosses show the BCGs without optical 
information, with the same colour code. Black circles show the galaxies
with radio emission (see Section 3.1 and Tables 1 and 2).}
\label{fig:fig1}
\end{figure}

Fig. \ref{fig:fig1} shows the absolute red magnitude for the 44 
objects in the sample with optical information available on the SDSS.
Those galaxies without information are plotted as crosses at a
fixed 
magnitude. The BCGs with radio emission are circled in black.
There seems to be no bias in redshift or cluster type for
those BCGs whose magnitude in unavailable on SDSS. Most of the BCGs 
with available information have absolute magnitude in the range 
--23 \simlt R \simlt --24, with few objects outside this interval.
The faintest objects are at the highest redshift in the sample and
are noticeably found in merging clusters.
The stellar masses for BCGs in the redshift range considered here are
on average a few times 10$^{11}$M$_{\rm Sun}$ (Lin et al. \cite{lin13}).

%
%

\subsection{Clusters with multiple BCGs}
The cluster sample consists of several dynamically disturbed clusters, 
some of them with multiple peaks in the X--ray surface brightness images. 
The identification of the BCG in those clusters was made under the
hypothesis that more than one BCG may be present, possibly associated
with merging sub--clusters. We considered only those BCGs falling within
the X--ray emission of the host cluster. Here below we report some
information. 

\begin{itemize}

\item{} A\,773 is a merging cluster with a radio halo
(Govoni et al. \cite{govoni01}). Two BCGs are located close to the single
X--ray peak. One has a compact radio source, detected after re--analyis of 
archival VLA observations at 1.4 GHz (Tables 1 and 2).

\item{} A\,2163 has two BCGs (Maurorgodato et al. \cite{maurogordato08}).
One of them is radio quiet, while the radio power of the second one is 
below the threshold considered for our statistical studies 
(Sect. 3.1, Tables 1 and 2). 

\item{} A\,2744 has two BGCs, located at the peak of both the
X--ray emission and the radio radio halo 
(V13, Giacintucci et al. in prep.). Both galaxies are radio quiet (Table 1).

\item{} RXCJ1314.4--2515 is a known merging system (Mazzotta et al. 
\cite{mazzotta11}) with two radio relics and a halo 
(V07, Feretti et al. \cite{feretti05}).
Two BCGs are identified by Valtchanov et al. (\cite{valtchanov02}).
They lie on each side of the single X--ray peak and are both radio--quiet
(Table 1).

\item{} RXCJ\,1514.9--1523 has two BGCs, both radio quiet.

\item{} Z\,5247 has a double peaked X--ray morphology  with a BCG at each peak. 
One of the two BCGs has detected radio emission (Table 1).

\end{itemize}

Only 3/10 of the BCGs in these multiple merger clusters show radio emission
at some level.

\subsection{Clusters with no BCGs}

In the optical field of A\,520, RXCJ\,2003.5--2323 and 
RXCJ\,1212.3--1816 there is no dominant galaxy that can be considered a BCG. 
It is noteworthy that they are all merging clusters, two of them with 
a radio halo (see Table 3).
\\
Three BCGs are actually reported in the literature for A\,520 (Mahdavi
et al. \cite{mahdavi07}), however they are all very distant from the 
centre of the X--ray emission, and their absolute optical magnitude is 
about a couple of magnitudes fainter than typical for this class of
objects, hence we regard A\,520 as a non--BCG cluster.

\subsection{Other special cases}

A\,141 shows complex X--ray substructure with a prominent
secondary peak south of the main double condensation. Based on the 
image inspection and on the literature information we considered
valid only the BCG listed in Hoffer et al. (\cite{hoffer12}). The
radio power of this galaxy is below the threshold considered for
our statistical investigations (see Sect. 3.1) and is reported in Table 2.

The central region of A\,2813 is quite complex, with three candidate 
galaxies within the brightest region of X--ray emission.
On the basis of the information available from NED (and by visual inspection
of DSS--2), the galaxy coincident with the peak of the X--ray emission is 
the faintest. The brightest galaxy in this region is 2\,MFGC~00530
and we consider it as the cluster BCG. It is radio quiet (see Table 1).

%
\begin{table*}[h!]
\caption[]{BCG Identifications. Radio powers and upper limits for the statistical sample }
Note: Dynamical state: M=merger, R=relaxed; 
Diffuse emission: mH=mini--halo; RH=Radio Halo; Rel=relic radio id. 
References to dynamical state: $^{\star}$ this paper; $^1$ Cuciti et al. 2015; 
$^2$ C10; $^3$ C13; $^4$ K15. 
References to diffuse emission: 
$^a$ Bacchi et al. 2003; $^b$V07; $^c$ K15; $^d$ K13; $^e$ Giacintucci et al.
2014: $^f$ Murgia et al. 2009; $^g$ V13: $^h$ Giacintucci et al. 2011b;
$^i$ Mazzotta \& Giacintucci 2008; $^l$ Giacintucci et al. 2014; 
$^m$ Orr\'u et al. 2007; $^n$ V08; $^O$ Venturi et al. 2011; 
$^p$ Feretti et al. 2005; $^q$ Giacintucci et al. 2011c; 
$^r$ Govoni et al. 2001; $^s$ Feretti et al. 2001.
\\  
\begin{tabular}{lllrcc}
\hline\noalign{\smallskip}
Name    & z & BCG ID  & logP$_{\rm 1.4~GHz}$ & Dyn. & Diff.   \\
        &  &2MASX/SDSS/other & W Hz$^{-1}$ & State & Emission \\

\hline\noalign{\smallskip}
A\,1763    & 0.2279 & 1237662306722447498 & 26.09 & M$^1$ &    \\ 
A\,2390    & 0.2329 & 1237680297268019748 & 25.54 & R$^2$ & mH$^a$ \\
S\,780     & 0.2357 & 2MASXJ14592875-1810453 & 25.21 & R$^2$ & mH$^{b,c}$ \\
RXCJ\,1115.8+0129 & 0.3499 & 1237654028716802393 & 24.87 & R$^2$ & \\
RXJ\,0439.0+0520 & 0.208 & 2MASXJ04390223+0520443 & 24.85 & R$^3$  & \\
RXJ\,1532.9+3021 & 0.345 & 2MASXJ15012308+4220405 & 24.80 & R$^2$ & mH$^{d,e}$ \\
A\,1835    & 0.252 & 2MASXJ14010204+0252423 & 24.80 & R$^1$ & mH$^f$ \\
A\,1576    & 0.302 & 2MASXJ12365866+6311145 & 24.77 & R$^3$ & \\
A\,1300    & 0.3075 & 2MASXJ11315413-1955391 & 24.77 & M$^2$ & RH$^g$ \\
RXCJ\,1504.1--0248 & 0.2153 & 1237655497600467190 & 24.76 & R$^2$ & mH$^h$ \\
RXJ2129.6+0005 & 0.235 & 2MASXJ21293995+0005207 & 24.64 & R$^4$ & mH$^c$ \\
A\,2667    & 0.2264 & 2MASXJ23513947-2605032 & 24.50 & R$^2$ & \\
Z\,5247    & 0.229 & 2MASXJ12342409+0947157 & 24.46 & M$^1$ & \\
A\,2146    & 0.234 & 2MASXJ15561395+6620530 & 24.44 & M$^3$ & \\
RXJ0027.6+2616 & 0.3649 & 2MASXJ00274579+2616264 & 24.38 & M$^3$ & \\
Z\,2701    & 0.214 & 2MASXJ09524915+5153053 & 24.32 & R$^2$ & \\
A\,1758a   & 0.28 & 2MASXJ13323845+5033351 & 24.27 & M$^2$ & RH$^g$ \\
Z\,2089    & 0.2347 & 2MASXJ09003684+2053402 & 24.25 & R$^2$ &  \\ 
RXJ2228.6+2037 & 0.4177 & 1237680298882433199 & $<$24.15 & M$^2$ & \\ 
Z\,2661    & 0.3825 & 1237667733956395341 & $<$24.06 & 
M$^{\star}$ & RH?$^n$ \\
A\,2261    & 0.224 & 2MASXJ17222717+3207571 & 24.04 & R$^2$ & \\
Z\,1953    & 0.373 & 2MASXJ08500730+3604203 & $<$ 24.04 & M$^4$ & \\ 
A\,2895    & 0.2275 & 2MASXJ01181108-2658122 & 24.02 & M$^1$ & \\
Z\,7160    & 0.2578 & 2MASXJ14571507+2220341 & 23.98 & R$^2$ & mH$^i$ \\
Z\,3146    & 0.29 & 2MASXJ10233960+0411116 & 23.95 & R$^4$ & mH$^{d,l}$ \\
A\,963     & 0.206 & 2MASXJ10170363+3902500 & 23.92 & R$^{\star}$ & \\
A\,1722    & 0.327 & ABELL1722:[HHP90]1318+7020A & $<$ 23.90 & R$^4$ & \\
Z\,348     & 0.254 & 1237666340799643767 & 23.89 & R$^3$ & \\
A\,2744    & 0.3066 & ABELL2744:[CN84]001 & $<$ 23.84 & M$^2$ & RH+Rel$^{g,m}$ \\
A\,2744\_1 & 0.3066 & ABELL2744:[CN84]002 & $<$23.84 & M$^2$ & RH+Rel$^{g,m}$ \\
Z\,5699    & 0.3063 & 2MASXJ13055884+2630487 & $<$23.84 & M$^3$ & \\
A\,781     & 0.2984 & 2MASXJ09202578+3029380 & $<$ 23.81 & M$^2$ & Rel$^{n,o}$\\
A\,2537    & 0.2966 & 2MASXJ23082221-0211315 & $<$ 23.80 & R$^2$ & \\
A\,2813    & 0.2924 & 2\,MFGC~00530 & $<$23.79 & M$^1$ & \\
Z\,7215    & 0.2897 & 2MASXJ15012308+4220405 & $<$ 23.78 & M$^{\star}$ & \\ 
A\,2631    & 0.2779 & 2MASXJ23373975+0016165 & $<$23.74 & M$^2$ & \\ 
RXJ0142.0+2131 & 0.28 & [BDJ2005]0479 & 23.72 & R$^3$ & \\
A\,1682    & 0.226 & 2MASXJ13064997+4633335 & 23.71 & M$^2$ & \\
RXCJ2211.7-0350 & 0.27 & 2MASXJ22114596-0349438 & $<$ 23.71 & R$^{\star}$ &  \\
Z\,5768    & 0.266 & 2MASXJ13114620+2201367 & $<$ 23.70 & M$^3$ & \\ 
A\,68      & 0.254 & 2MASXJ00370686+0909236 & $<$ 23.65 & M$^1$ & \\ 
A\,2645    & 0.251 & 2MASXJ23411705-0901110 & $<$23.64 & M$^3$ & \\
A\,2485    & 0.2472 & 2MASXJ22483112-1606258 & $<$23.62 & R$^3$ & \\ 
RXCJ\,1314.4-2515 & 0.2439 & 2MASXJ13142209-2515456 & $<$23.61 & M & 
RH+2 Rel$^{b,p}$ \\
                  &        & 2MASXJ13143263-2515266 & $<$23.61 & & \\
A\,2697    & 0.232 & 2MASXJ00031162-0605305 & $<$ 23.56 & R$^3$ & \\ 
RXCJ0437.1+0043 & 0.2842 & 2MASXJ04370955+0043533 & 23.55 & R$^2$ & \\
A\,3444    & 0.2542 & 2MASXJ10235019-2715232 & 23.55 & R$^1$ & mH$^c$ \\
A\,267     & 0.23 & 2MASXJ01524199+0100257 & $<$23.55 & M$^2$ & \\ 
%
%
Z\,5247\_1 & 0.229 & 2MASXJ12341746+0945577 & $<$23.55 & M$^1$  & \\
A\,2111    & 0.229 & 2MASXJ15394049+3425276 & $<$23.55 & M$^{\star}$ & \\ 
A\,2219 & 0.2281 & 2MASXJ16401981+4642409 & $<$23.54 & M$^2$ & RH$^m$ \\
RXCJ\,1514.9-1523\_1 & 0.2226 & 2MASXJ15145772-1523447 & $<$23.52 & M$^1$ 
& RH$^q$ \\
RXCJ\,1514.9--1523 & 0.2226 & 2MASXJ15150305-1521537 & $<$23.52 & M$^1$ 
& RH$^q$ \\
RXCJ\,0510.7--0801 & 0.2195 & 2MASXJ05104786-0801449 & $<$23.51 & M$^1$ & \\
A\,773     & 0.217 & 2MASXJ09175344+5143379 & $<$23.49 & M$^2$ & RH$^r$ \\
A\,1423    & 0.213 & 2MASXJ11571737+3336399 & $<$23.48 & R$^2$ & \\
A\,209     & 0.206 & 2MASXJ01315250-1336409 & $<$23.44 & M$^2$ & RH$^b$ \\
A\,2163    & 0.203 & ABELL2163:[MCF2008]308 & $<$23.43 & M$^2$ & RH$^s$ \\
%
\hline\noalign{\smallskip}
\end{tabular}
\label{tab:id}
\end{table*}
%
%
%
%
\begin{table*}[h!]
\caption[]{BCG Identifications. Radio powers and upper limits for the ``faint'' sample.}
Note: Dynamical state: M=merger, R=relaxed; 
Diffuse emission: mH=mini--halo; RH=Radio Halo; Rel=relic radio id. 
References to dynamical state: $^1$ Cuciti et al. 2015; $^2$ C10;
$^3$ C13. References to diffuse emission: $^a$ V08; $^b$ Macario et al. 2010;
$^c$ Govoni et al. 2001; $^d$ Feretti et al. 2001; $^e$ Brunetti et al. 2008.
\\  
\begin{tabular}{lllrcc}
\hline\noalign{\smallskip}
Name    & z & BCG ID  & logP$_{\rm 1.4~GHz}$ & Dyn. & Diff. \\
        &  &2MASX/SDSS/other & W Hz$^{-1}$ & State & Emission \\
\hline\noalign{\smallskip}
A\,2552    & 0.301 & 2MASXJ23113330+0338056 & 23.37 & R?$^1$ & \\
A\,697     & 0.282 & 2MASXJ08425763+3622000 & 23.30 & M$^2$ & RH$^{a,b}$ \\
A\,141  & 0.23 & 2MASXJ01053543-2437476 & 23.21 & M$^2$ & \\
A\,773\_1  & 0.217 & 2MASXJ09175344+5144009 & 23.21 & M$^2$ & RH$^{c}$ \\
A\,611     & 0.288 & 2MASXJ08005684+3603234 & 23.12 & R$^2$  & \\
RXJ\,0439.0+0715 & 0.244 & 2MASXJ04390053+0716038 & 23.12 & R$^3$ & \\
A\,2163\_1 & 0.203 & 2MASX J16153353-0609167 & 22.99 & M$^2$ & RH$^d$ \\
A\,3088    & 0.2537 & 2MASXJ03070207-2839574 & 22.79 & R$^2$ & \\
A\,521     & 0.2475 & 2MASXJ04540687-1013247 & 22.66 & M$^2$ & RH+Rel$^e$ \\
\hline\noalign{\smallskip}
\end{tabular}
\label{tab:id2}
\end{table*}
%
%

\begin{table}[h!]
\caption[]{Clusters without BCG}
References to dynamical state: $^1$ C10; $^2$ K15. 
$^{\star}$ Note on the presence of diffuse radio emission: 
GRH=giant radio halo. References: $^a$ Giacintucci et al. 2009;
$^b$ Govoni et al. 2001. \\
\begin{tabular}{lccc}
\hline\hline\noalign{\smallskip}
Name    & z & Dynamical State & Note$^{\star}$   \\
        &   &                 &                \\
\hline\noalign{\smallskip}
RXCJ\,2003.5--2323 & 0.317  &  M$^1$ & GRH$^a$\\
RXCJ\,1212.3--1816 & 0.269  &  M$^2$ & --     \\
A\,520            & 0.203   &  M$^1$ & GRH$^b$\\
\hline\noalign{\smallskip}
\end{tabular}
\label{tab:noBCG}
\end{table}

%

\section{The radio and X--ray data}\label{sec:radioX}

\subsection{The radio data}

The EGRHS is the starting point of our BCG sample. The radio information
we used to derive the radio luminosity function, however, is not taken 
from the 610 MHz GMRT observations (V07, V08, K13 and K15).
To ensure a sensitivity as uniform as possible over the whole sample, and 
for a direct comparison with works from other authors, mainly performed 
at 1.4 GHz, we cross--checked our sample with the Northern VLA Sky Survey 
(NVSS) and extracted the 1.4 GHz flux density of each source directly from 
those images. To overcome a few cases of blending on NVSS, where possible 
we used the images on the VLA FIRST Survey (angular resolution of 
5$^{\prime\prime}$). Finally, for those BCGs in mini--halo clusters (see
Table 1) we used the 1.4 GHz flux density values published in
Giacintucci et al. (\cite{giacintucci14}), which were accurately estimated
to avoid contamination from the diffuse emission of the 
mini--halo\footnote{For the BCG in S\,780, flux density measurements from 
GMRT proprietary data and re--analysis of VLA archival data suggest that 
the source is variable, and that the core of the radio emission has an 
inverted spectrum.}.

For a number of clusters the FIRST images, whose angular resolution is
comparable to our 610 MHz GMRT images ($\sim 5^{\prime\prime}$) are not 
available, and the angular resolution of the NVSS 
($45^{\prime\prime}\times45^{\prime\prime}$) is inadequate to isolate the flux 
density of the
BCG from that of other nearby sources (RXCJ\,0439.0+0520, RXCJ\,0027.6+2616, 
Z\,348, RXCJ\,0142.0+2131 and A\,1682) or from a diffuse radio halo 
(A\,1300 and A\,1758a).
In those cases we made use of our high resolution images at 610 MHz
(V07, V08, K13 and K15) and at 325 MHz 
(Venturi et al. \cite{venturi13}) and  we derived the flux density at 1.4 GHz 
assuming a spectral index of 0.8, which is a reasonable average value
for this type of sources (Klein et al. \cite{klein95}).

Due to the higher sensitivity of the EGRHS compared to NVSS and FIRST, 
six BCGs detected at 610 MHz do not have a counterpart either on
NVSS or on FIRST. In particular: 
A\,141 and A\,3088 (V07), 
A\,521 (Giacintucci et al. \cite{giacintucci06}),
A\,697 (V08), 
A\,2552 (K15) and RXJ\,0439.0+0715 (K13).
Moreover, radio emission below the sensitivity limit of FIRST and NVSS 
was detected by Giacintucci et al. (in prep.) for 
A\,773\_1 and A\,2163\_1 after re-analysis of archival 1.4 GHz VLA
data.
Finally, A\,611 has a strong detection at 610 MHz (the radio source
associated with the BCG has S$_{\rm 610~MHz}=59\pm3$ mJy), but nothing is visible 
on NVSS. Inspection of FIRST shows a very weak source (at $\sim4\sigma$) 
which would have remained unnoticed without careful comparison with the 
610 MHz image.  For this source, in Table 2 we report the radio power 
derived from the 1.4 GHz flux density from FIRST. 
These nine radio sources, listed in Table 2, were removed from our statistical 
analysis.
\\
For those BCGs without radio emission, we considered a conservative
radio power upper limit derived from NVSS, whose average noise level is 
1$\sigma$=0.45 mJy~b$^{-1}$, i.e. S$_{\rm 1.4~GHz} \le 2.25$ mJy 
(namely $5\sigma$, see Table \ref{tab:id}).

The final sample we used for the statistical studies (see 
Sect. 4 and 5) includes a total of 59 BCGs, hosted in 55 clusters.

The histogram in Fig. 2 shows the radio power distribution 
of all BCGs with radio emission, including those nine faintest objects 
(see Table 2) which we did not consider in the statistical analyis 
performed in Section 4.
The distribution peaks around logP$_{\rm 1.4~GHz} (W Hz^{-1}) \sim$ 24.5, 
which is the typical transition power between FRI and FRII radio galaxies 
(Fanaroff \& Riley \cite{fr74}), as commonly found at cluster centres. 
Except for the case of the BCG in A\,1763 (the only object in the bin of
highest radio power), the BCGs in relaxed clusters are the most abundant
in most bins of radio power. 

As a final remark, we note that all the radio BCGs in our sample show very 
little extended structure both at 610 MHz and at 1.4 GHz. The only exception is 
the wide--angle tail (WAT) at the centre of A\,1763, whose 1.4 GHz contours 
from FIRST are shown in Fig. 3 overlaid on the optical frame and on the 
$Chandra$ X--ray emission. This is also the most powerful radio source in 
the sample (see Table 1). WAT radio galaxies are found only in association 
with brightest cluster galaxies (Feretti \& Venturi \cite{fv02} for a review),
and their bent morphology is considered as the signature of bulk motions in 
the intracluster medium (Burns \cite{burns98}). 


\begin{figure}[htbp]
\centering
\includegraphics[angle=0,scale=0.43]{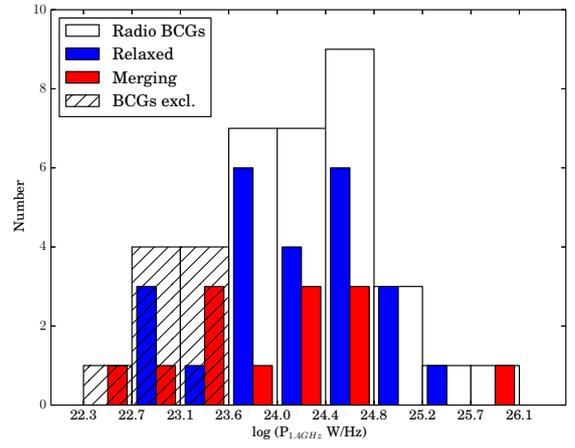}
\caption{Histogram of the radio power of the BCGs in the sample. The shaded 
intervals refer to those BCGs which were removed from the statistical 
analysis (Sect. 3.1).}
\label{fig:fig2}
\end{figure}
%

%
%
\begin{figure*}[htbp]
\includegraphics[angle=0,scale=0.46]{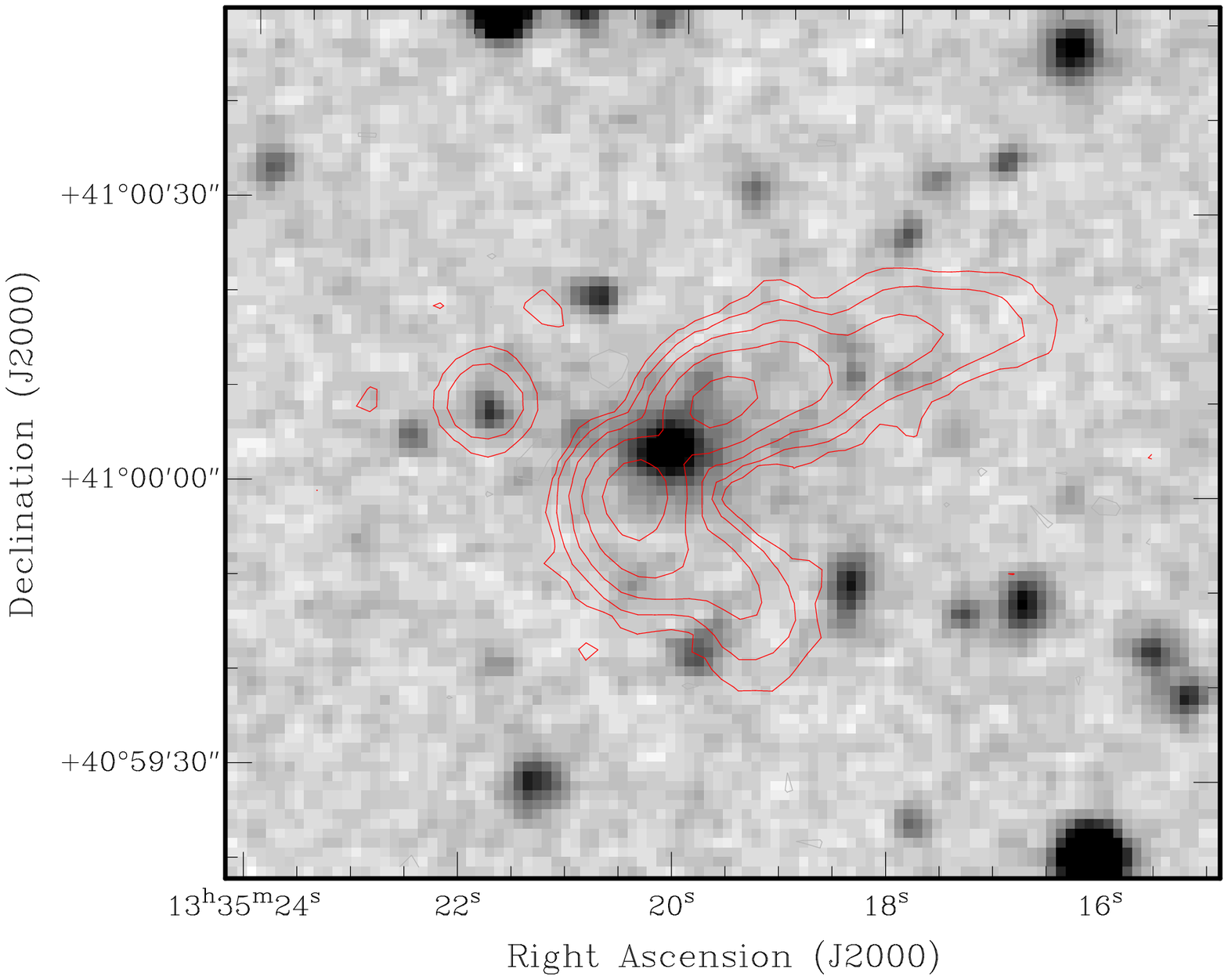}
\hspace{0.5truecm}
\includegraphics[angle=0,scale=0.52]{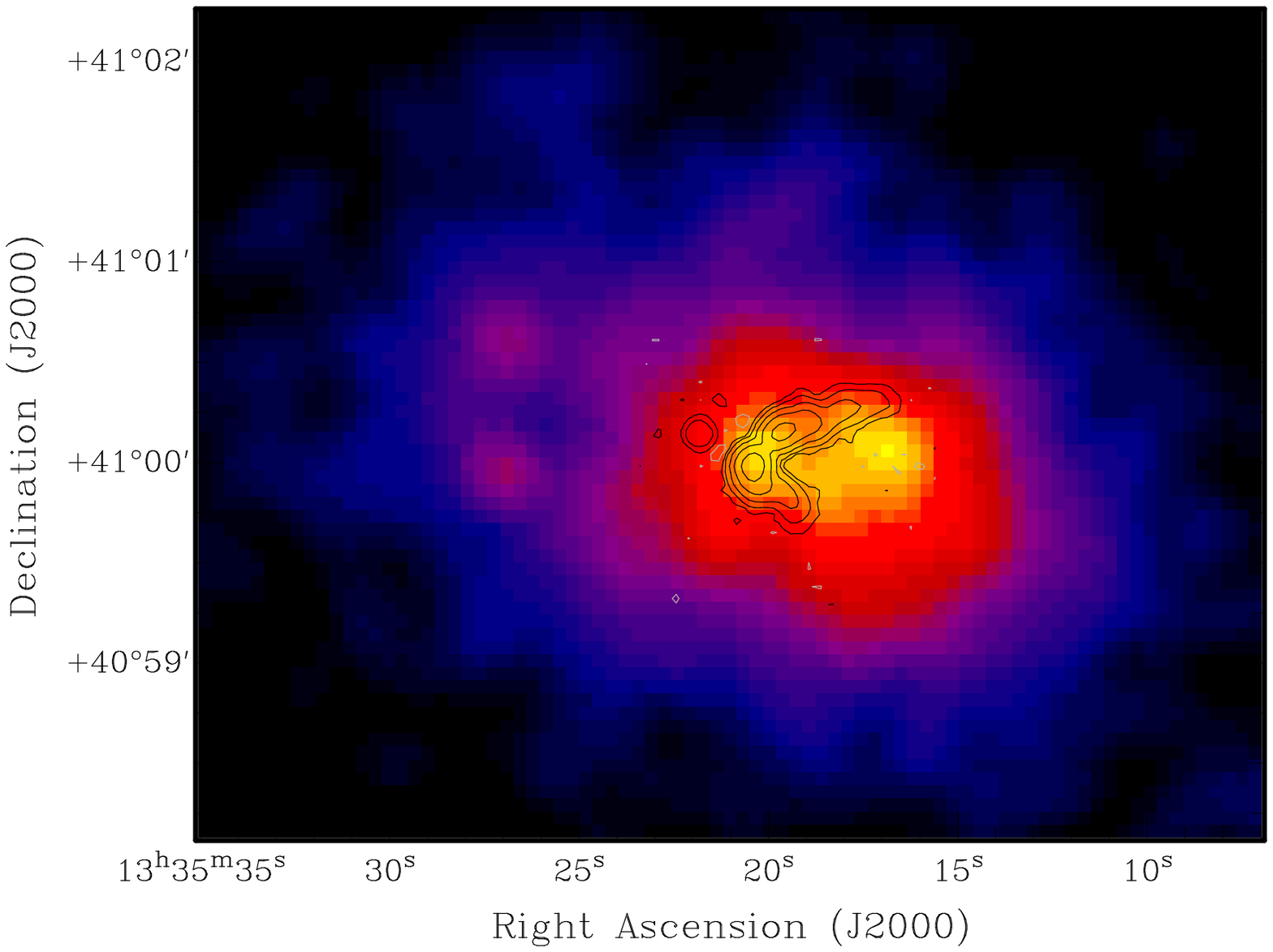}
\caption{1.4 GHz contours from FIRST of the wide--angle tail in A\,1763 
overlaid on the red optical frame of the ESO Digitized Sky Survey DSS2 
(left panel) and on the {\it Chandra} image (right panel).
Radio contours are: $\pm$0.4, 1.6, 6.4, 25.6, 102.4 mJy/b in both panels
(negative contours are shown in grey).}
\label{fig:fig3}
\end{figure*}
%

\subsection{The X--ray data and cluster morphological analysis}

To assess the cluster dynamical status we used the morphological parameters,
namely the power ratio P$_3$/P$_0$, the centroid shift $w_{\rm 500}$ and the 
concentration parameter $c_{\rm 100}$, derived in C10 and Cassano et al. 
(\cite{cassano13}, hereinafter C13) from proprietary and archival
$Chandra$ data.
 
We briefly remind here that the power ratio is a multipole
decomposition of the two--dimensional projected mass within a given aperture, 
and it provides a measure of the substructure (e.g., Buote \& Tsai 
\cite{bt95}). The centroid shift $w$ is defined as the standard deviation 
of the projected separation between the peak and the centroid of the cluster 
X--ray brightness distribution (e.g., Poole et al. \cite{poole06}). In
particular, $w_{\rm 500}$ is estimated over an aperture of 500 kpc.
The concentration parameter $c_{\rm 100}$ is defined as the ratio of the peak 
(within 100 kpc) over the ambient (within 500 kpc) X--ray surface brightness 
(Santos et al. \cite{santos08}).

For five clusters not included in those earlier works 
(marked with $\star$ in Table 1) we derived the morphological indicators 
following C10 (see Sect. 3 of C10 for details).
The results of this analysis are reported in  Tables \ref{tab:id}, 2 
and \ref{tab:noBCG}, where the clusters are classified as merger (M) 
or relaxed (R), according to their position in the morphological
diagrams. 
 
We further complemented our analysis with literature information, by visual 
inspection of the available X--ray images, and considering the presence
of diffuse extended emission in the form of radio halos or mini--halos,
whose connection respectively with merging and relaxed clusters is an
established result (Brunetti \& Jones \cite{bj14}).

All 65 clusters in the EGRHS have a classification of their dynamical 
status: 35 are merging (M) and 30 are relaxed (R), i.e.
54\% and 46\% respectively (see Tables 1, 2 and 3). 
Note that the morphological parameters used to derive the cluster dynamics 
are not sensitive to mergers aligned close to the line of sight,
however it is reasonable to assume that these are only a negligible fraction
of the whole esample.
The redshift distribution of the two subsamples is quite similar: the 
median value of z is 0.251 and 0.253 for the merging and relaxed clusters
respectively.

\section{Radio loudness, cluster dynamics and radio luminosity 
function}\label{sec:analysis}

%
\begin{figure*}[htbp]
\centering
\includegraphics[angle=0,scale=0.38]{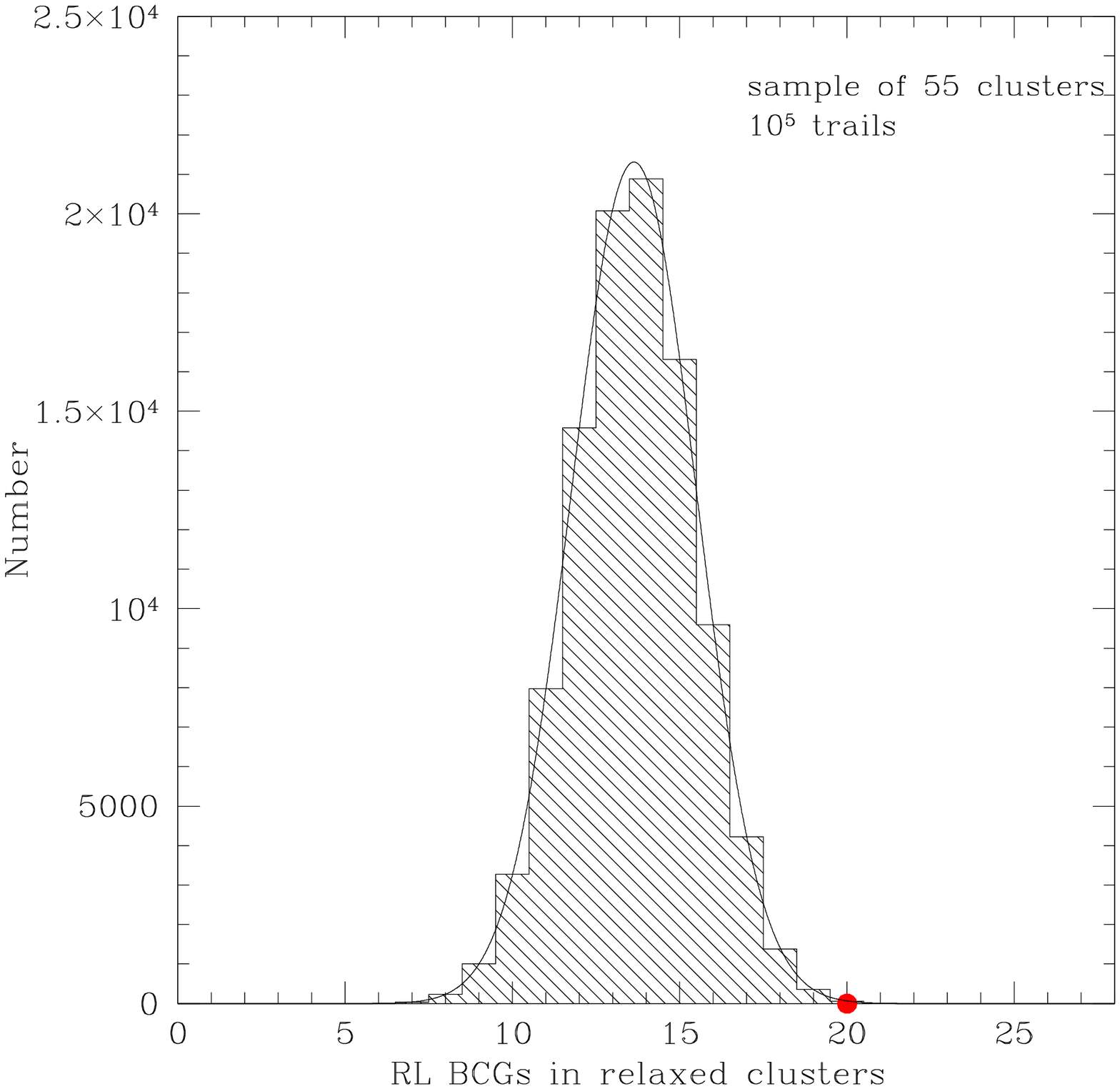}
\hspace{0.4truecm}
\includegraphics[angle=0,scale=0.47]{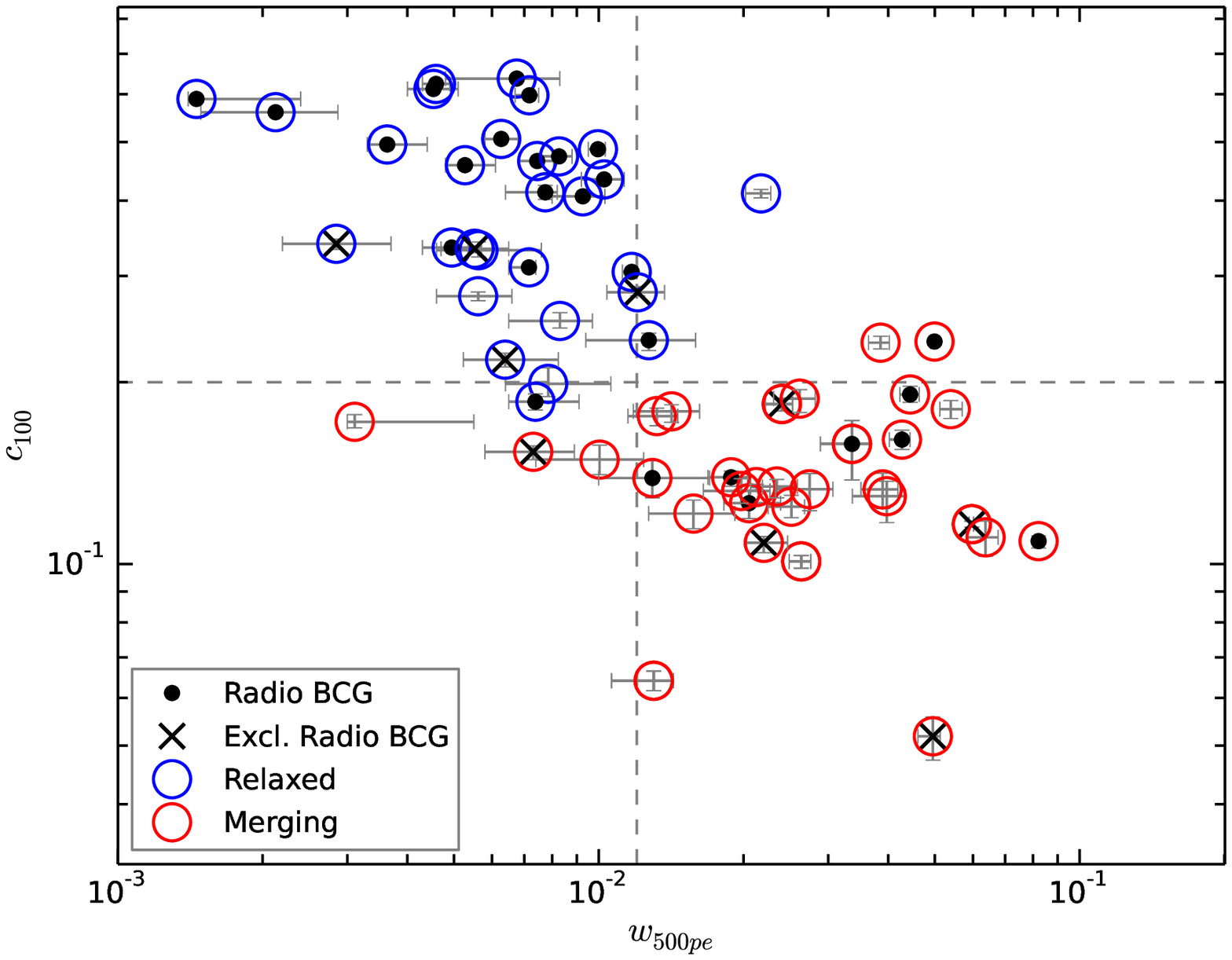}
\caption{
{\it Left Panel:} Result of the Monte Carlo calculations. The histogram
shows the distribution of the radio loud BCGs in relaxed clusters after 
$10^5$ trials. The red dots shows the 3.4$\sigma$ location of our observed
result compared to the random distribution.
{\it Right panel:} Distribution of the BCGs in the 
$w_{\rm 500}$--$c_{\rm 100}$ space. Relaxed clusters are shown as blue circles, 
and are all located in the upper left  quadrant; merging clusters are 
shown as red cicles, and occupy the lower right portion. Filled black 
circles are the  radio loud BCGs, the black crosses show the BCGs with radio 
emission which have been removed from the analysis (Sect. 3.1).
The threshold values to classify clusters as mergers are as in C10, i.e., 
P$_3$/P$_0~>~1.2\times10^{-7}$, $w_{\rm 500} > 0.012$ and $c_{\rm 100}~<~0.20$.}
\label{fig:fig4}
\end{figure*}

%

\subsection{Radio loudness fraction and cluster dynamics}

Starting from the initial sample of 68 BCGs, and after inspection of
the radio information available (Sect. 3.1), we ended up with a final 
sample of 59 objects for our statistical analysis.
Out of these 59 BCGs, 28 are radio loud and 31 are radio quiet 
(47\% and 53\% in each class, see Table \ref{tab:id}). 

We divided the 55 clusters hosting the 59 BCGs (see Sect. 2.1 for the
clusters with multiple BCGs) in merging and relaxed, following Sect. 3.2 
(see column 5 in Table 1), to estimate the fraction of radio loud BCGs  in 
each class: we have 29 merging and 26 relaxed clusters (53\% and 47\% 
of the total respectively).
Our results are summarised in Table 4. 
Radio loud BCGs are considerably more common in relaxed clusters (71\%
against 29\% in merging clusters). Conversely, radio quiet BCGs are much
more common in merging systems (81\% to be compared to the 19\% in 
relaxed clusters).
\\
We tested the significance of this result by running Monte Carlo (MC)
calculations. In particular, we randomly assigned the 28 radio loud BCGs 
among the 55 clusters in the sample and counted the number of objects 
that fall in relaxed clusters in our MC trials. In the left panel of
Fig. 4 we report the distribution of the number of radio loud BCGs
in relaxed clusters obtained after $10^5$ MC trials. The distribution
can be fitted with a Gaussian function, with a central value of 13.6 and
standard deviation of 1.875. This means that the observed value of 20 BCGs 
in relaxed clusters (red point in Fig. 4, left panel) is at 3.4$\sigma$ 
from the value expected 
assuming that the distribution of the radio loud BCGs is independent of 
the cluster dynamical status. This shows that the probability that our
result is a chance detection is $\le~3.4\times10^{-4}$. 
A similar result can be obtained considering radio quiet BCGs in merging 
and relaxed clusters.

The distribution of the BCGs in the different environments is given in the 
right panel of Fig. 4, which shows
%
%
the clusters in the $w_{\rm 500}$--$c_{\rm 100}$ space, selected to describe 
the dynamical state. The grey dotted lines are traced as reference values 
to statistically pinpoint the regions of merging (bottom right portion) 
and relaxed (upper left portion) clusters (see C10 for details).
The black points show the radio loud BCGs, and again we note that the bulk
of them (71\%) are found in relaxed systems.

\subsection{Radio power of BCGs and cluster dynamics}

The radio power of the BCGs in the full sample of 68 objects spans more 
than three orders of magnitude, from logP$_{\rm 1.4~GHz}$(W Hz$^{-1}$)=22.66 
(A\,521) to logP$_{\rm 1.4~GHz}$ (W Hz$^{-1}$)=26.09 (A\,1763) (see Tables 1
and 2). Even restricting our considerations to the 59 BCGs used in the 
statistical sample, the range of values is quite broad (2.5 orders of 
magnitude).

We checked for a possible connection between the radio power of the BCGs,
the X--ray luminosity (L$_{\rm X}$),  
and the dynamical state of the host cluster. The left panel of Fig. 5 
shows the distribution of the radio power of the BCGs in different
environments versus L$_{\rm X}$.  Upper limits are also plotted.
The figure is suggestive of a few considerations. Even though no significant
trend is visible, upper limits are much more abundant in less
luminous (logL$_{\rm X}$ (erg~s$^{-1}) \le 45.1$ ) merging clusters,
whereas for logL$_{\rm X}$ (erg~s$^{-1}) > 45.1$  the fraction of radio BCGs is 
much higher, and 9 out of 10 are found in relaxed clusters.

If we consider the BCGs in relaxed clusters, only a weak trend is
present between the radio power and the core X--ray luminosity (within 
0.15R$_{500}$\footnote{R$_{500}$ is the radius corresponding to a total
density contrast 500$\rho_c$(z), where $\rho_c$(z) is the critical density.},
see C13) of the  host cluster: a Spearman test on a possible correlation
between these two variables provides $\rho$=0.375 and a probability
of null hypothesis of 16.8\%. A correlation might be present for
logL$_{\rm X}$ (erg~s$^{-1}$) \simgt 45, but the small number of points
above this value does not allow to draw any conclusion.

Finally, we checked for a possible dependence of the BCG radio power
with the cluster dynamics. We used the concentration parameter $c_{100}$ as
proxy for the cluster dynamical state 
(high values of $c_{100}$ indicate peaked X--ray brightness distributions, 
typical of relaxed clusters) and plotted the radio power and upper
limits vs $c_{100}$. Our results are shown in the right panel of Fig. 5. 
As a further information, we highlighted those relaxed clusters hosting 
a mini--halo, and those merging clusters hosting a giant radio halo.
By definition all relaxed clusters (blue dots) have $c_{\rm 100}\ge0.2$.
\\
The radio power range is populated fairly uniformly for both classes. 
No obvious trend is visible for the radio loud BCGs in merging clusters,
which span the whole range of radio power. On the other hand, the
right panel of Fig. 5 is suggestive of a positive trend for the BCGs in 
relaxed systems, which show increasing radio power with increasing value 
of $c_{\rm 100}$.
A Spearman test on the possible correlation between logP$_{\rm 1.4~GHz}$ and 
$c_{\rm 100}$ for the radio BCGs in relaxed clusters provides a Spearman's 
rank--order coefficient $\rho$=0.53 (suggesting a positive correlation)
and a probability of null hypothesis of 1\%, which is thus rejected.
The test was performed on all relaxed clusters with $c_{\rm 100} > 0.2$.
The inclusion of the two most deviating points (i.e., A\,1576 and A\,2390)
does not change the result. 
 
%
\begin{table}[h!]
\caption[]{Radio loudness fraction in merging and relaxed clusters}
\begin{tabular}{lrrcc}
\hline\noalign{\smallskip}
\# BCGs    & Merging & Relaxed & \% Merging & \% Relaxed  \\
           &                  &                  &            &             \\
\hline\noalign{\smallskip}
Radio Loud  & 8  & 20 & 29\% & 71\% \\
Radio Quiet & 25 & 6 & 81\% & 19\% \\
\hline\noalign{\smallskip}
\end{tabular}
\label{tab:fractions}
\end{table}
%

%
\begin{figure*}[htbp]

\includegraphics[angle=0,scale=0.45]{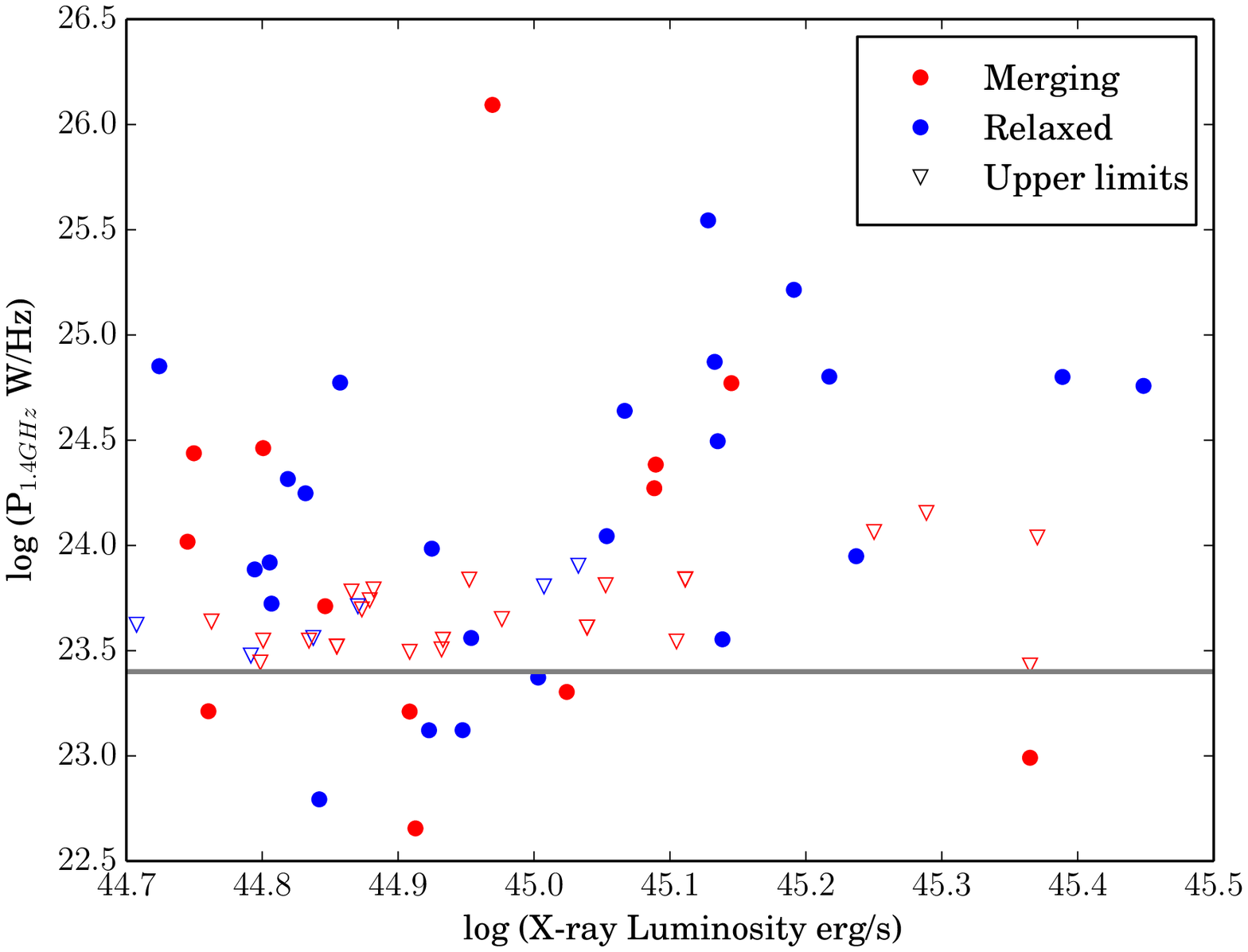}
\hspace{0.6truecm}
\includegraphics[angle=0,scale=0.45]{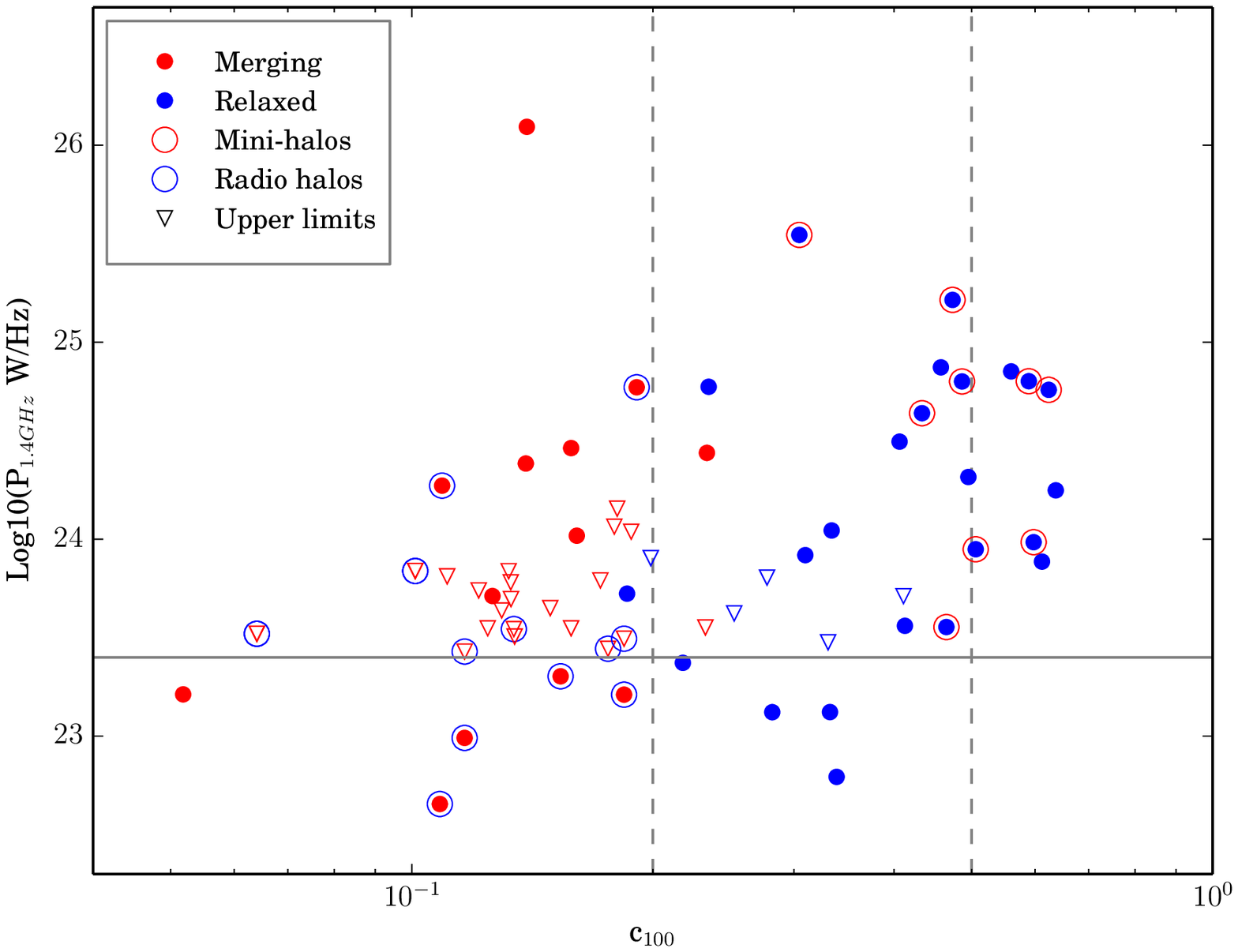}
\caption{
{\it Left panel:} Distribution of the logL$_{\rm X}$--logP$_{\rm 1.4~GHz}$
for the sample, where  logL$_{\rm X}$ is the X--ray luminosity of the 
host cluster. Undetected BCGs are shown as triangles (blue for
relaxed, red for merging clusters). The objects below the grey line are 
those undetected in NVSS and hence removed from the statistical study.
{\it Right Panel:}
Distribution of the radio power of the BCGs in the sample
as a function of the concentration parameter of the hosting cluster
($c_{\rm 100}$). Merging clusters are shown as red dots, relaxed clusters
are blue circles. Upper limits in each class are shown as triangles,
with the same colour code. The information on the presence of mini--halos
in relaxed systems and radio halos in merging clusters is also shown.
The grey horizontal line is the same as in the left panel. Dashed 
vertical lines are drawn at $c_{\rm 100}$=0.2 and 0.5 (see Sect. 5.1).}
\label{fig:fig5}
\end{figure*}
%

\subsection{The fractional radio luminosity function}

The fractional radio luminosity function (RLF) is a powerful tool to 
investigate the statistical properties of a population of objects. 
It provides the probability that a galaxy is radio loud with 
radio power higher than a given value of P.

To minimize the problems raised by the sensitivity limits of different 
arrays and ensure uniform sensitivity, we based our analysis on NVSS 
(see Section 3.1). However, the redshift of the BCGs in our sample spans 
over the range z=0.2--0.4, and the radio power upper limit 
for the undetected sources is a function of redshift, as clear from 
Table 1.
To account for the upper limits in the fractional RLF, different methods have 
been proposed. Among those, we adopted that developed by Fanti 
(see appendix in Hummel \cite{hummel81}), which shows the smallest 
statistical fluctuations in Monte Carlo experiments. The cumulative 
fractional radio luminosity function $F(\ge P_k)$ is described as follows:
\\
$$F(\ge P_k)=\sum_{j=1}^{k} f(P_j)$$
\\
with 
\\
$$f(P_k)={{1-\sum_{j=1}^{k-1} f(P_j)}\over{n_u(P_i<P_k)+n_d(P \le P_k)}}.$$
\\
Here $f(P_j)$ is the fraction of detections in the $j$--th radio power 
interval, 
$n_u (P_i<P_k)$ is the number of upper limits (undetections) for 
$P_i<P_k$, and $n_d(P \le P_k)$ is the number of detections for $P\le P_k$.
Given a sample with N objects,
due to sensitivity limits $n_u$ objects are undetected, and $n_d$ are
detected, and $n_u$+$n_d$=N.  

We computed the fractional RLF using radio power intervals with width 
$\Delta$logP$_{\rm 1.4~GHz}$=0.4. We then summed up the detections in 
each bin to obtain the RLF in the cumulative form. Our results are 
shown in Fig. 6, where the RLF is reported for the merging 
(red) and relaxed clusters (blue).

The fractional RLF for our two subsamples is different. 
BCGs in relaxed clusters show a significantly higher probability to be 
radio loud than those in merging systems. At high radio powers the statistics 
is poor (as clear from Tab. 1), while the faint end of the radio 
luminosity function suffers from incompleteness, nevertheless the 
differences are clear in the most populated bins of radio 
detections.
In particular, the probability that a BCG in a relaxed cluster is radio loud 
with radio power logP$_{\rm 1.4~GHz}$(W~Hz$^{-1}$) \simgt 23.5 is $\sim$90\%, 
to be compared to $\sim$30\% for the BCGs in merging clusters.

To quantify this result we applied a Kolmogorov-Smirnov (KS) test to the 
fractional RLF in merging and relaxed clusters. The null hypothesis in this
test is that both samples are drawn from the same distribution. We obtained
D=0.67 (distribution parameter) and p=7.6\% (probability function), hence we
reject the null hypothesis.
The statistical significance of this result is small, however it is 
strengthened by the results found in the previous sections.

%
\begin{figure}[htbp]
\includegraphics[angle=0,scale=0.45]{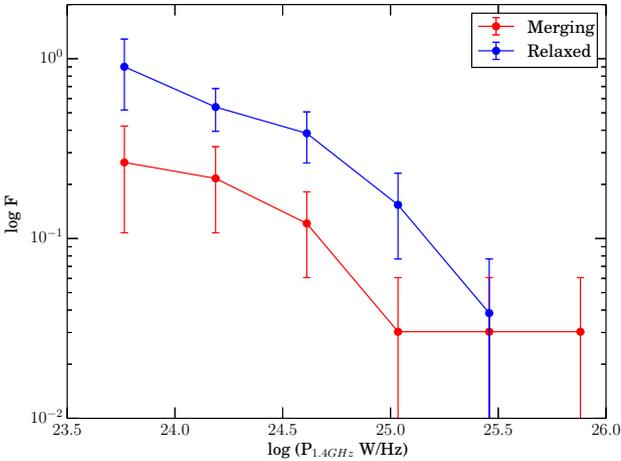}
\caption{Cumulative Radio Luminosity Function for the BCGs in merging 
(red) and relaxed clusters (blue). F is defined in Sect. 4.3.
The points are plotted in the middle value of each bin.}
\label{fig:fig6}
\end{figure}
%

\section{Discussion}\label{sec:disc}

In this paper we address the question of the possible connection between the 
nuclear radio emission of BCGs and the dynamical state of the host cluster. 
The sample selected to this aim consists of 68 galaxies in the clusters 
of the Extended GMRT Radio Halo Survey 
(L$_X$ (0.1-2.4 keV) $> 5\times10^{44}$ erg s$^{-1}$, $0.2 \le z \le 0.4 $
and $\delta  > -31^{\circ}$).

Previous studies focussed on samples of BCGs in nearby clusters and groups 
(z$<$0.2), and were mainly performed in the framework of the AGN/ICM feedback 
mechanisms, hence they were mainly concentrated on the connection between 
the radio loudness of the BCGs and  the presence of a cool core (Mittal et al. 
\cite{mittal09}, Sun et al. \cite{sun09}, Dunn \& Fabian \cite{dunn08}). 
They provided clear indication of a positive correlation between the strength 
of the cluster cool core and the presence of a radio loud BCG, and its power. 

The statistical analysis presented here differs from the earlier
investigations, and complements those results. Our BCG sample covers
a new redshift space, and extends the findings of previous studies up to z=0.4.
Our starting sample of galaxy clusters has no a-priori bias towards 
the cluster dynamical state, the clusters being almost evenly distributed 
between mergers and relaxed systems (52\% and 48\% respectively). 
From the radio point of view, our analysis is not restricted to the most 
powerful and extreme BCGs, but it
includes also relatively faint objects, down to 
logP$_{\rm 1.4~GHz}$ (W~Hz$^{-1}$)$\sim$ 23.5. 
Finally, our statistical analysis accounts for the sensitivity limits of 
the radio data, with the inclusions of the upper limits for the 
non--detections, and it is hence complete within the volume covered by
the sample.

\subsection{Statistical properties of the BCGs in the EGRHS cluster sample and 
the cluster dynamics}

The radio power range of our sample is typical of low and intermediate power 
radio loud AGN (23 \simlt logP$_{\rm 1.4~GHz}$W~Hz$^{-1}$ \simlt 26, 
see Owen \& Laing \cite{ol89} for reference values), and our 
analysis shows that the nuclear radio properties of BCGs strongly 
depend on the central properties of the host cluster. 

Roughly 50\% of the BCGs in our sample are radio loud, consistent with 
earlier findings (Best et al. \cite{best07}). However, if we separate the BCGs
according to the dynamical state of the cluster, we find that radio loud  
BCGs are much more abundant in relaxed systems, i.e. 71\% vs 29\%. Monte
Carlo simulations show that the probability of a chance result can be 
rejected (Sect. 4.1). At the same time radio quiet objects are mainly found 
in merging clusters, i.e. 81\% vs 19\%. 
Motivated by this result, we further investigated the possible dependence 
of the radio  power of the BCGs with the cluster dynamics. 

Mittal et al. (\cite{mittal09}), reported that at fixed X--ray luminosity
the fraction of radio loud BCGs as well as their radio power increase 
with increasing cool--core strength. Our analysis 
based on the X--ray images for the clusters in our sample neither separates 
cool--core and non--cool--core clusters, nor allows it to measure the 
strength of cool cores; however, the parameter $c_{100}$ was introduced
to search for cool--core clusters 
and its value is a measure of the cool--core strength, which increases 
with increasing $c_{100}$ (Santos et al. \cite{santos08}, C13).
On this basis, we made a comparison between our results and those  
in Mittal et al. (\cite{mittal09}), and tentatively separated all clusters in 
three different intervals of $c_{100}$, following their classification 
(see dotted lines in the right panel of Fig. 5): clusters with 
$c_{100} < 0.2$ are mergers, i.e. likely non--cool--core clusters (NCC), 
those with $0.2 < c_{100} < 0.5$ are weak cool--cores (WCC) and those with 
$c_{100} > 0.5$ should have strong cool--cores (SCC). Allowing for the 
uncertainty
due to the poorer statistics in each $c_{100}$ interval, our data are consistent
with the result reported in Mittal et al. (\cite{mittal09}).
In particular, the fraction of radio BCGs is considerably lower in clusters 
with $c_{100} < 0.2$ (NCC) compared to WCC+SCC clusters ($c_{100} > 0.2$): 
it rises from  $\sim$33\% to $\sim$85\%. 
If we restrict our considerations to the WCC and SCC clusters, the fraction 
further increases from $\sim$84\% for WCC to 100\% for SCC, as clear in 
Fig. 5 (right).

Finally, we found a positive correlation between the concentration parameter
and the radio power of the BCG in relaxed clusters. After removing  
A\,1576 and A\,2390 (see Sect. 4.1), we fitted the law 
$${\rm logP}_{\rm 10}{\rm (1.4~GHz)=a~log}_{10} c_{\rm 100}+{\rm b}$$ 
and found a=2.55$\pm$0.33 and b=25.05$\pm$0.83. 
For comparison, Mittal et al. (\cite{mittal09}) found 
P$_{1.4~GHz} \propto t_{\rm cool}^{-3.16\pm0.38}$, on a larger sample of objects.
Considering that $t_{cool} \propto c_{100}^{-1}$ (e.g. Santos et al.
\cite{santos08}) our results are in line with those obtained by 
Mittal et al. (\cite{mittal09}).

The close connection between the fraction of radio loud BCGs, their power
and the cluster dynamics is further strengthened by the fractional radio
luminosity function (RLF), which shows that BCGs in relaxed systems have a 
considerably higher probability to be radio loud than those in merging
clusters: the probability to find a radio loud BCG with 
logP$_{\rm 1.4~GHz}$W~Hz$^{-1} \simgt$ 23.5 is almost 90\% for relaxed clusters,
whereas it falls to $\sim$ 20-30\% in merging clusters.

\subsection{Optical properties of the BCGs in the EGRH sample}

To check for possible trends between the dynamical state of the host clusters, 
the radio emission of the BCGs and the properties of the gaseous environment 
feeding the AGN, we investigated the occurrence of  optical emission lines 
in the spectra of the BGCs in our sample.

Unfortunately, only 28/68 BCGs in the full sample have spectra on the 
SDSS, therefore no strong conclusions can be drawn. Nevertheless the
results are interesting. 
Among those 28 BCGs, 19 show radio emission (5/19 in merging and 14/19 
in relaxed clusters) and 9 have upper limits (7/9 in merging and 2/9 in 
relaxed clusters). 
None of the galaxies with radio upper limit shows emission lines in the 
optical spectrum, while 11/19 radio loud do, and they are all found 
in relaxed clusters. 

The optical spectrum of the BCG is available only for 7 of the mini--halo 
clusters in our sample , and in all cases emission
lines are present. Those are A\,1835, A\,2390, RXCJ\,1504.1--0248,
RXCJ\,1532.9+3021, RXCJ\,2129.6+0005, Z\,3146 and Z\,7160. The remaining
four are RXCJ\,1115.8+0129, Z\,2089, Z\,2701 and Z\,348.

It is interesting to note that the radio power of these emission line 
galaxies in our sample is typical of low/intermediate power 
radio galaxies, which usually lack emission lines (Balmaverde et al. 
\cite{balmaverde08}, Heckman \& Best \cite{hb14} and references therein) 
Moreover, most of the radio emitting BCGs in the sample are unresolved. 
The only exceptions are the small tail in Z\,5247 
and the large WAT in A\,1763 (see Fig. 3).

\subsection{BCGs, accretion and the radio properties of the host clusters} 

The dynamical state of galaxy clusters and their overall formation 
through a series of mergers are phenomena involving scales much larger (Mpc) 
than the inner regions at play in the radio loud activity of AGN, whose
typical sizes are of the order of the sub--kpc. However, the striking 
occurrence of radio--loud BCGs in relaxed clusters and the positive trend 
between radio power and cool--core strength suggest that the two phenomena 
are related, at least to some extent.

The radio emission in massive early--type galaxies is broadly classified 
on the basis of their accretion rate, with ``jet--mode'' radio galaxies 
accreting inefficiently (L/L$_{\rm Edd}$ \simlt 0.01) and ``radiative--mode''
sources accreting at high rates, i.e. L/L$_{\rm Edd}$ \simgt 0.1 (see 
Heckman \& Best \cite{hb14} for a recent review and nomenclature). 
The two classes of radio-loud AGN differ in the optical properties, the 
former being low-excitation (LERG), the latter being high excitation 
galaxies (HERG). Typically, radio galaxies with moderate radio power 
(P$_{\rm 1.4~GHz}$ \simlt $10^{25}$ W~Hz$^{-1}$) belong to the first class 
(e.g., Balmaverde et al. \cite{balmaverde08}), while more powerful  
radio galaxies (P$_{\rm 1.4~GHz}$ \simgt $10^{25}$ W~Hz$^{-1}$) usually show 
spectral features typical of the HERG class.
It has been recently proposed (Hardcastle et al. \cite{hardcastle07}) 
that a main difference between LERG and HERG resides in the source of 
accreting gas: LERG may accrete hot gas from the intergalactic medium
(``hot--mode''), while HERG may be fed by infalling cold gas (``cold--mode'').
It has been further pointed out that the radio luminosity function for 
``hot--mode''  radio galaxies would depend both on the black hole mass 
function and on the distribution properties of the central hot gas, while 
that of the ``cold--mode'' radio galaxies would be not be affected by the 
black hole mass. 

By definition, cool--core clusters are characterized by relatively cold gas 
with high mass deposition rates at their centres (see Hudson et al. 
\cite{hudson10} for a review). This builds up a reservoir of cold gas for 
the BCGs at their centres, which may feed the AGN and provide fuel to the 
radio emission. Though on a limited fraction of our sample, the available
spectral information does support this, suggesting that at least part of 
the radio loud BCGs in cool--core clusters in our sample are supplied 
by cold gas at the cluster centre. An impressive example is the 
BCG in RXCJ\,1504.1--0248 (Ogrean et al. \cite{ogrean10}), which is one 
of the cool--core clusters with mini--halo. 
At the same time, the observations show that cool cores are often disrupted 
during cluster mergers, and this cold gas may no longer be available to the 
BCGs in unrelaxed systems. Indeed, none of the emission line BCGs in our
sub--sample (see previous section) is in a merging cluster.

It is tempting to suggest that the fraction of radio galaxies and the 
fractional radio luminosity function for the BCGs in the relaxed clusters of 
our sample is the result of two populations of radio galaxies, one
accreting gas from the hot corona, and the other accreting cold gas in
cluster core region. The latter would not be found in merging clusters, 
as the contribution of the intracluster cold gas would be missing.

\section{Summary and Conclusions}\label{sec:conc}
In this paper we addressed the possible connection between the radio 
properties of Brightest Cluster Galaxies and the dynamical state of 
the host cluster using a sample of BCGs selected from the Extended GMRT 
Radio Halo Sample (EGRHS). The BCGs in our sample are located in the 
redshift interval 0.2$\le$z$\le$0.4. All clusters have available quantitative 
information on their dynamical state from $Chandra$ X--ray data. 
We can summarize our results as follows:

\begin{itemize}

\item{} Most of the BCGs in our sample have optical red magnitude 
in the range --24 \simlt R \simlt --23, i.e. the stellar mass range of the 
galaxies (a few times 10$^{11}$M$_{\rm Sun}$) is narrow. Considering
that the fraction of radio galaxies and the radio luminosity function
depend on the stellar mass (e.g., Auriemma et al. \cite{auriemma77}, 
Ledlow \& Owen \cite{lo96}, Bardelli et al. \cite{bardelli10}), the optical 
properties of our sample ensure that our results are not strongly 
affected by this. 

\item{} The full radio sample includes 68 BCGs, whose radio power spans 
a wide range, from
logP$_{\rm 1.4~GHz}$ (W~Hz$^{-1}$)=22.8 to 26.1. Most of the radio galaxies
are unresolved at the resolution of few arcsec (GMRT at 610 MHz).

\item{} High quality X--ray imaging is available for all the 65 clusters
in the EGRHS. Our quantitative morphological analysis shows that mergers 
and relaxed clusters are fairly equally represented, i.e. 54\% and 
46\% respectively.

\item{} Among the sample of 59 BCGs considered for our statistical
analysis, 47\% is radio loud. Among the radio--loud population 71\% of 
the BCGs are located in relaxed systems. This result is solid ($3.4\sigma$): 
Monte Carlo simulations  show that the probability that this result is 
a chance detection is $\le 3.4\times10^{-4}$. On the other hand, radio quiet 
BCGs are mostly found in merging systems (81\%). 

\item{} We find that the fraction of radio BCGs in relaxed clusters increases 
with increasing value of the concentration parameter $c_{\rm 100}$, reaching 
100\% for $c_{\rm 100}>0.5$. For relaxed clusters ($c_{\rm 100} > 0.2$),
we find a positive trend between $c_{\rm 100}$ and the BCG radio power, in 
the form P$_{\rm 1.4~GHz}\propto c_{\rm 100}^{2.55\pm0.33}$.
Since $c_{\rm 100}$ is an indicator of the cool--core strength 
($c_{\rm 100} \propto$t$_{\rm cool}^{-1}$, e.g. Santos et al. \cite{santos08}),
this trend suggests that the most powerful BCGs are located in the 
strongest cool--core clusters, indicating a clear connection between the
AGN activity of the BCGs and the deposition of the cooling gas at the
cluster centre (see also Mittal et al. \cite{mittal09}).

\item{} For the BCGs in relaxed clusters, there is only a weak correlation 
between the radio power of the BCGs and the core X--ray luminosity of the 
host cluster (within 0.15R$_{500}$).

\item{} The fractional radio luminosity function differs for the BCGs in
the two environments. In particular, the BCGs in relaxed clusters have 
an extremely high probability to be radio loud, i.e. almost 90\%,  
to be compared to the $\sim$20--30\% for those in relaxed clusters.

\item{} For a subset of our full sample ($\sim$41\%), optical
spectra are available in the SDSS. 11/28 of those spectra show 
emission lines, and these are all radio--loud BCGs in relaxed clusters,
7 of them with a radio mini--halo.

\end{itemize}

Our study provides support for a strong link between the radio properties 
of BCGs and the dynamical state of the host cluster.
We propose that our results reflect the AGN accretion mode of the
BCGs. At least a fraction of the radio loud BCGs in relaxed clusters may 
be accreting cold gas from the central region of the host cool--core cluster. 
Such cold gas is certainly available for those radio galaxies in 
mini--halo clusters, as their optical spectra show, and it is most likely 
less abundant in BCGs in merging clusters, where the dominant accretion mode 
for the radio AGN may be due to accretion of hot gas from the IGM of the 
galaxy itself.
Understanding how the cold gas in the central cluster regions is transported
all the way through the galaxy in the nearest proximity of the black hole 
remains an open issue.

\begin{acknowledgements}

We thank Prof. R. Fanti for the many insightful discussions.
Thanks are due to S. Ettori for providing the routines we used for the 
$Chandra$ X--ray data analysis and for the derivation of the cluster 
morphological parameters.
R.K. and T.V. acknowledge partial support by PRIN-INAF 2008 and by  
FP7-People-2009 IRSES CAFEGroups project under grant agreement 247653. 
Funding for SDSS-III has been provided by the Alfred 
P. Sloan Foundation, the Participating Institutions, the National Science 
Foundation, and the U.S. Department of Energy Office of Science. The SDSS-III 
web site is http://www.sdss3.org/.
SDSS-III is managed by the Astrophysical Research Consortium for the 
Participating Institutions of the SDSS-III Collaboration including the 
University of Arizona, the Brazilian Participation Group, Brookhaven National 
Laboratory, Carnegie Mellon University, University of Florida, the French 
Participation Group, the German Participation Group, Harvard University, the 
Instituto de Astrofisica de Canarias, the Michigan State/Notre Dame/JINA 
Participation Group, Johns Hopkins University, Lawrence Berkeley National 
Laboratory, Max Planck Institute for Astrophysics, Max Planck Institute for 
Extraterrestrial Physics, New Mexico State University, New York University, 
Ohio State University, Pennsylvania State University, University of Portsmouth,
Princeton University, the Spanish Participation Group, University of Tokyo, 
University of Utah, Vanderbilt University, University of Virginia, University 
of Washington, and Yale University. 
This research has made use of the NASA/IPAC Extragalactic Database (NED) 
which is operated by the Jet Propulsion Laboratory, California Institute of 
Technology, under contract with the National Aeronautics and Space 
Administration. 

\end{acknowledgements}

\end{document}